\documentclass[twocolumn,showpacs,preprintnumbers,amsmath,amssymb]{revtex4}
\usepackage{graphicx}
\usepackage{dcolumn}
\usepackage{bm}
\begin{document}
\title{Spin gap in the 2D electron system of GaAs/AlGaAs single
heterojunctions\\ in weak magnetic fields}
\author{V.~S. Khrapai, A.~A. Shashkin, and E.~L. Shangina}
\affiliation{Institute of Solid State Physics, Chernogolovka,
Moscow District 142432, Russia}
\author{V. Pellegrini and F. Beltram}
\affiliation{NEST-INFM, Scuola Normale Superiore, Piazza dei
Cavalieri 7, I-56126 Pisa, Italy}
\author{G. Biasiol and L. Sorba$^*$}
\affiliation{NEST-INFM and Laboratorio Nazionale TASC-INFM, I-34012
Trieste, Italy\\
$^*$Universita di Modena e Reggio Emilia, I-41100 Modena, Italy}
\begin{abstract}
We study the interaction-enhanced spin gaps in the two-dimensional
electron gas confined in GaAs/AlGaAs single heterojunctions subjected
to weak magnetic fields. The values are obtained from the chemical
potential jumps measured by magnetocapacitance. The gap increase with
parallel magnetic field indicates that the lowest-lying charged
excitations are accompanied with a single spin flip at the
odd-integer filling factor $\nu=1$ and $\nu=3$, in disagreement with
the concept of skyrmions.
\end{abstract}
\maketitle

\section{INTRODUCTION}

Much interest has been attracted recently by possible formation of
the spin textures in two-dimensional (2D) electron systems in
perpendicular magnetic fields. The spin gap in a 2D electron system
is expected to be enhanced compared to the single-particle Zeeman
energy due to electron-electron interactions \cite{ando,sondhi}.
Within the concept of exchange-enhanced gaps, the gap enhancement is
given by the exchange energy of a single spin-flip excitation
\cite{ando}. However, in the limit of low single-particle Zeeman
energies (particularly, at weak magnetic fields) and of weak
electron-electron interactions, the skyrmion --- the spin texture
characterized by many flipped spins --- is predicted to become for
filling factor $\nu=1$ the lowest-energy charge-carrying excitation
\cite{sondhi}. This corresponds to a reduction of the
exchange-enhanced gap.

Possible formation of the skyrmions remains controversial so far
although much work has been done on this (see, e.g.,
Refs.~\cite{usher,schmeller,maude,barrett,goldberg,qq,zhitomir,terasawa}).
Strongly enhanced values of the spin gap in the 2D electron system in
GaAs/AlGaAs heterostructures were found by measurements of the
activation energy for the longitudinal resistivity minimum at filling
factor $\nu=1$, 3, and 5 \cite{usher,schmeller,maude}. It was argued
in Ref.~\cite{schmeller} that introducing a parallel component of the
magnetic field allows determination of the spin of the lowest-lying
charged excitations by the increase in the activation energy with
parallel field. The experimental results indicated that the
excitations at $\nu=1$ are accompanied with seven electron spin
flips, while at $\nu=3$ and 5 only a single spin flips
\cite{schmeller,terasawa}. This is consistent, in principle, with the
predictions of the skyrmion approach. However, one should be careful
in interpreting results of activation energy measurements because
they yield a mobility gap which may be different from the gap in the
spectrum. The latter can be determined directly by measurements of
the chemical potential jump across the gap based on
magnetocapacitance spectroscopy \cite{smith,aristov}.

In this paper, we report the first measurements of the chemical
potential jump across the many-body enhanced spin gap at $\nu=1$ and
3 in the 2D electron system in GaAs/AlGaAs single heterojunctions in
weak magnetic fields using a magnetocapacitance technique. We find
that the increase of the gap with parallel component of the magnetic
field corresponds to a single spin flip for both $\nu=1$ and $\nu=3$,
which does not support formation of the skyrmions in the range of
magnetic fields studied, down to $B_\perp\approx2$~T. This finding is
in contrast to the results of indirect transport measurements
\cite{schmeller} on very similar samples in the same range of
magnetic fields. Concerning the observed slightly-sublinear
dependence of the spin gap on perpendicular magnetic field, we
suggest that its origin can be related to the Landau level mixing due
to electron-electron interactions.

The remainder of the paper is organized as follows. Thermodynamic
measurement technique and samples are described in Sec.~\ref{exp}.
Details of the data analysis and experimental results on the behavior
of the spin gap with magnetic field are given in Sec.~\ref{res}. The
obtained results are discussed and compared to those of transport
measurements in Sec.~\ref{disc}. The main results are summarized in
the conclusion.

\section{EXPERIMENTAL TECHNIQUE}
\label{exp}

\begin{figure}
\scalebox{0.55}{\includegraphics[clip]{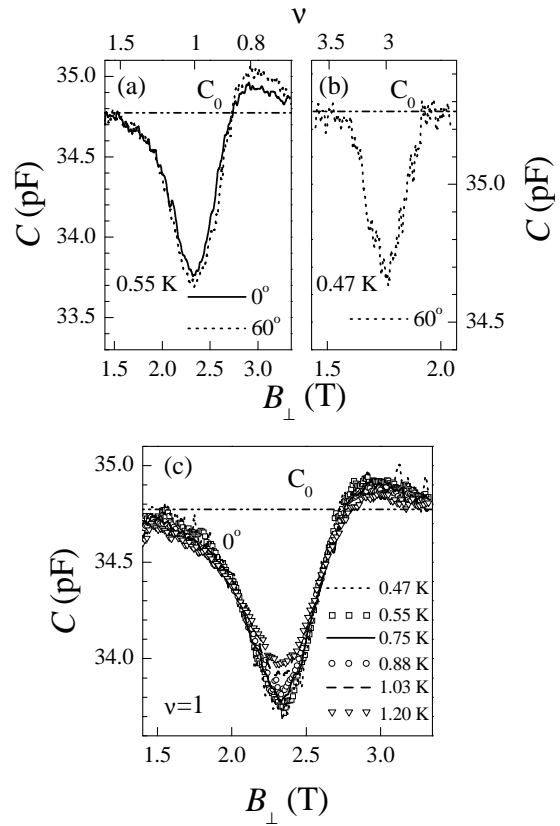}}
\caption{\label{fig1} Magnetocapacitance traces in the low
temperature limit at different tilt angles for $n_s=5.7\times
10^{10}$~cm$^{-2}$ (a) and $n_s=1.28\times 10^{11}$~cm$^{-2}$ (b) and
the temperature dependence of the magnetocapacitance at
$n_s=5.7\times 10^{10}$~cm$^{-2}$ (c). Also shown by a dash-dotted
line is the geometric capacitance $C_0$.}
\end{figure}

Measurements were made in an Oxford dilution refrigerator with a base
temperature of $\approx30$~mK on remotely doped GaAs/AlGaAs single
heterojunctions (with a low temperature mobility $\approx2\times
10^6$~cm$^2$/Vs at electron density $1\times 10^{11}$~cm$^{-2}$)
having the Hall bar geometry with area $5\times 10^4$~$\mu$m$^2$. A
metallic gate was deposited onto the surface of the sample, which
allowed variation of the electron density by applying a dc bias
between the gate and the 2D electrons. To populate the 2D electron
system, the sample was illuminated with an infrared light-emitting
diode; after the electron density saturated, the diode was switched
off. The gate voltage was modulated with a small ac voltage of 4~mV
at frequencies in the range 0.5--11~Hz, and both the imaginary and
real components of the current were measured using a current-voltage
converter and a lock-in amplifier. Smallness of the real current
component as well as proportionality of the imaginary current
component to the excitation frequency ensure that we reach the low
frequency limit and the measured magnetocapacitance is not distorted
by lateral transport effects. A dip in the magnetocapacitance at
integer filling factor is directly related to a jump of the chemical
potential across a corresponding gap in the spectrum of the 2D
electron system \cite{smith}:
\begin{equation}\frac{1}{C}=\frac{1}{C_0}+\frac{1}{Ae^2dn_s/d\mu},\label{C}\end{equation}
where $C_0$ is the geometric capacitance between the gate and the 2D
electrons, $A$ is the sample area, and the derivative $dn_s/d\mu$ of
the electron density over the chemical potential is thermodynamic
density of states. The chemical potential jump is determined by
integrating the magnetocapacitance over the dip in the low
temperature limit where the magnetocapacitance saturates and becomes
independent of temperature \cite{gapsi} (see Fig.~\ref{fig1}).
Additional experiments were performed on three-electrode samples of
GaAs/AlGaAs single heterojunctions in which the 2D electron system is
field-effect induced in a way similar to silicon
metal-oxide-semiconductor field-effect transistors. It is separated
from the front gate by a blocking barrier and from the back electrode
by a wide but shallow tunnel barrier. The sample design allows the
suppression of lateral transport effects, so the range of strong
perpendicular magnetic fields becomes easily accessible (for more
details, see Ref.~\cite{aristov}).

\section{RESULTS}
\label{res}

Typical magnetocapacitance traces in the low temperature limit at
different electron densities and tilt angles of the magnetic field as
well as the temperature dependence of the magnetocapacitance are
displayed in Fig.~\ref{fig1} near $\nu=1$ and 3. Narrow minima in the
magnetocapacitance at integer filling factor are separated by broad
maxima, the oscillation pattern reflecting the behavior of the
thermodynamic density of states in quantizing magnetic fields. As the
magnetic field is increased, the maximum $C(B)$ increases and
approaches in the high-field limit the geometric capacitance $C_0$.
We have verified that the obtained $C_0$ corresponds to the value
calculated using Eq.~(\ref{C}) from the zero-field capacitance and
the density of states $m/\pi\hbar^2$ (where $m=0.067m_e$ and $m_e$ is
the free electron mass). Note that the so-called negative
compressibility effect manifests itself in our samples as a local
maximum in $C(B)$ above $C_0$ that is observed in fields
$B_\perp>2$~T at the edge of the dip in the magnetocapacitance for
$\nu=1$. Experimentally, it is easier to analyze $C(B)$ traces: being
independent of $B_\perp$, the geometric capacitance $C_0$ practically
does not depend on parallel component of the magnetic field but
increases with $n_s$ as the 2D electrons are forced closer to the
interface. As explained above, the chemical potential jump at integer
$\nu=\nu_0$ is determined by the area of the dip in the
magnetocapacitance:
\begin{equation}\Delta=\frac{Ae^3\nu_0}{hcC_0}\int_{\text{dip}}\frac{C_0-C}{C}dB_\perp,\label{Delta}\end{equation}
where the integration over $B_\perp$ is equivalent to the one over
$n_s$ provided the minimum is narrow. The criterion of narrow minima
is met in our experiment. Indeed, the formula~(\ref{Delta}) gives
values of $\Delta$ that are underestimated approximately by $(\delta
B/B)^2$, where $\delta B/B$ is the relative half-width of the nearly
symmetric minimum. This contribution is less than 4\% even in the
lowest magnetic fields $B_\perp$ used in the experiment. More
importantly, it does not depend on parallel component of the magnetic
field (Fig.~\ref{fig1}(a)) and, therefore, the data analysis made
below is valid.

\begin{figure}
\scalebox{0.4}{\includegraphics[clip]{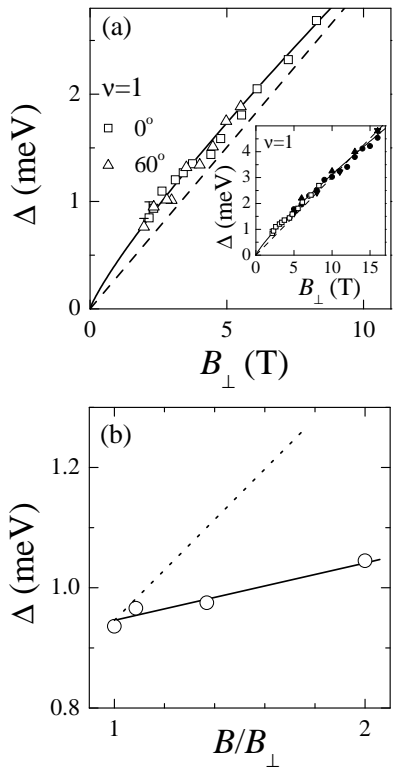}}
\caption{\label{fig2} (a) Chemical potential jump across the spin gap
at $\nu=1$ as a function of perpendicular component of the magnetic
field excluding the term that is responsible for the increase of the
gap with parallel field. The level width contribution is indicated by
systematic error bars; see text. The solid line is a power-law fit
with exponent $\alpha\approx0.85$, and the dashed line corresponds to
$g=5.2$. The data in perpendicular magnetic fields are compared in
the inset with those of Ref.~\cite{aristov} obtained on
three-electrode samples (solid symbols). (b) Change of the $\nu=1$
spin gap with $B_\parallel$ at fixed $B_\perp=2.35$~T. The solid line
corresponds to an effective $g$ factor $g\approx0.7$. The dashed line
depicts the slope expected from Ref.~\cite{schmeller}.}
\end{figure}

In magnetic fields $B_\perp\lesssim2$~T, where the magnetocapacitance
does not reach $C_0$, the value $C_0$ in the integrand of
Eq.~(\ref{Delta}) is replaced by a step function $C_{\text{ref}}$
that is defined by two reference levels corresponding to the $C$
values at $\nu=\nu_0+1/2$ and $\nu=\nu_0-1/2$. The so-determined
$\Delta$ is smaller than the level splitting by the level width whose
contribution is obtained by substituting
$(C_0-C_{\text{ref}})B_0^2/CB_\perp^2$ (where $B_0=hcn_s/e\nu_0$) for
the integrand in Eq.~(\ref{Delta}) and integrating between the
magnetic fields $B_1=hcn_s/e(\nu_0+1/2)$ and
$B_2=hcn_s/e(\nu_0-1/2)$. It is clear that in the range of $B_\perp$
mentioned above, the accuracy of the measurement method becomes worse
with decreasing $B_\perp$ due to increasing level broadening/overlap.

In Fig.~\ref{fig2}(a), we show the chemical potential jump across the
$\nu=1$ spin gap as a function of perpendicular component of the
magnetic field. In the high-field limit the data are in good
agreement with the results obtained on three-electrode samples (see
the inset to Fig.~\ref{fig2}(a)). This indicates that the enhanced
value of gap is sample independent. As $B_\perp$ is decreased, the
$g$ factor $g=g_{\text{max}}$ becomes yet more enhanced than its
high-field value $g\approx5.2$ (dashed line). This implies that in
the weak-field region, the functional form of $\Delta(B_\perp)$
changes to a sublinear law. Particularly, the data for the gap can be
described by a function $\Delta\propto B_\perp^\alpha$ with
$\alpha\approx0.85$ (solid line). Once the exponent $\alpha$ is close
to unity, this sublinear fit is practically indistinguishable from
the linear fit to the data in fields $B_\perp\gtrsim5$~T.

\begin{figure}
\scalebox{0.4}{\includegraphics[clip]{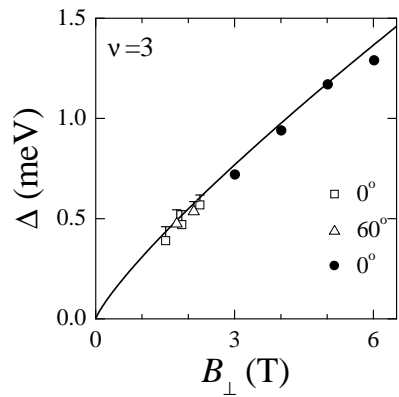}}
\caption{\label{fig3} The spin gap at $\nu=3$ vs $B_\perp$ excluding
the same term $g_{\text{min}}\mu_B(B-B_\perp)$ as for the case of
$\nu=1$ (open symbols) along with the data obtained on
three-electrode samples (solid symbols). The level width contribution
is shown by systematic error bars. The solid line is a power-law fit
with exponent $\alpha\approx0.85$.}
\end{figure}

In Fig.~\ref{fig2}(b), we show the behavior of the spin gap at
$\nu=1$ in a fixed field $B_\perp=2.35$~T as a function of parallel
magnetic field. The gap increases linearly with total magnetic field
$B$ and is described by an effective $g$ factor
$g=g_{\text{min}}\approx0.7$. Note that this value is considerably
smaller than the effective $g$ factor $g\approx3.1$ (the dashed line
in Fig.~\ref{fig2}(b)) obtained by transport measurements on very
similar samples in the same range of magnetic fields
\cite{schmeller}. Moreover, the determined $g_{\text{min}}\approx0.7$
is confirmed by the data of Fig.~\ref{fig2}(a), where we compare the
results for the $\nu=1$ spin gap in perpendicular and tilted magnetic
fields excluding the term $g_{\text{min}}\mu_B(B-B_\perp)$ that
describes the increase of the spin gap with $B_\parallel$. The data
coincidence indicates that the value $g_{\text{min}}$ does not
practically change in the range of fields $B_\perp$ (or electron
densities) studied. That stands to reason that due to weak dependence
of the gap on parallel magnetic field, the accuracy of the method for
determining $g_{\text{min}}$ is not high (50\% at worst for our
case). Nevertheless, this is not crucial for our results and
conclusions.

The chemical potential jump across the $\nu=3$ spin gap versus
perpendicular component of the magnetic field is displayed in
Fig.~\ref{fig3}. The gap can also be described by a power-law
dependence $B_\perp^\alpha$ with $\alpha\approx0.85$, although its
value is about 30\% smaller than that at $\nu=1$. The same term
$g_{\text{min}}\mu_B(B-B_\perp)$ as for the case of $\nu=1$ has been
subtracted from the value of gap in tilted magnetic fields. As
inferred from the coincidence of the data in perpendicular and tilted
magnetic fields, there is no pronounced dependence of
$g_{\text{min}}$ on filling factor.

\section{DISCUSSION}
\label{disc}

It is tempting to compare the obtained $g_{\text{min}}\approx0.7$
with the ``non-interacting'' value $|g|\approx0.44$, which is
determined in spin resonance measurement \cite{dobers}, as dictated
by Kohn's theorem analog. However, for magnetotransport and
magnetocapacitance experiments, it is the interaction-enhanced values
of $g$ that are relevant. Recently, it has been established in
studies of weak-field Shubnikov-de~Haas oscillations and
parallel-field magnetotransport \cite{zhu} that at low electron
densities, the $g$ factor is enhanced well above its value $|g|=0.44$
in bulk GaAs, being renormalized by electron-electron interactions.
The $g$ factor $g=g_{\text{min}}$ determined in our experiment turns
out to be concurrent with the interaction-renormalized $g$ factor
($g=0.7$ at electron density $4\times10^{10}$~cm$^{-2}$) obtained in
Ref.~\cite{zhu}. We therefore arrive at a conclusion that the
lowest-lying charged excitations are accompanied with a single spin
flip at $\nu=1$ and 3.

While the results of indirect transport studies of the spin gap in
the 2D electron system in GaAs have been interpreted as evidence for
skyrmions \cite{schmeller}, our direct measurements of the spin gap
in very similar samples question such an interpretation. The obtained
experimental results do not support formation of the skyrmions in the
range of magnetic fields studied, down to $B_\perp\approx2$~T. As a
matter of fact, the change of the spin gap with parallel magnetic
field is consistent with the concept of exchange-enhanced gaps which
includes single spin-flip excitations, the Zeeman energy contribution
being determined by the interaction-renormalized $g$ factor.
Therefore, evidence for skyrmions should be sought in the region of
yet lower Zeeman energies, using direct experimental methods.

We now discuss briefly the Landau level mixing as a possible
candidate to explain the experimental dependence of the spin gap on
perpendicular magnetic field. The gap enhancement is determined by
the exchange energy estimated as $e^2/\kappa l$ (where $l$ is the
magnetic length), which yields a square-root magnetic field
dependence of the gap if $g_{\text{max}}\gg g_{\text{min}}$. For our
case the corrections to the exchange energy due to level overlap
\cite{gumbs} are below 5\% in magnetic fields $B_\perp$ about 2~T
and, therefore, they cannot provide an appreciable increase of the
power of the theoretical square-root dependence $\Delta(B_\perp)$
\cite{aristov}. We suggest that the Landau level mixing due to
electron-electron interactions \cite{gumbs,mihalek,falko}, which
gives rise to a more pronounced reduction of the gap in low magnetic
fields, can be responsible for the observed dependence
$\Delta(B_\perp)$.

\begin{figure}
\scalebox{0.4}{\includegraphics[clip]{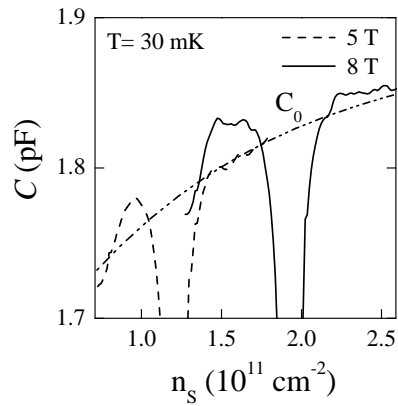}}
\caption{\label{fig4} Capacitance near $\nu=1$ as a function of
electron density in three-electrode samples at different
perpendicular magnetic fields. The geometric capacitance $C_0$ is
indicated by the dash-dotted line.}
\end{figure}

Since mixing the Landau levels causes electron-hole symmetry
breaking, its significance can easily be established in experiment.
Evidence for the electron-hole symmetry breaking is given by the data
for the magnetocapacitance which reveals the negative compressibility
effect at $\nu=1$ (Fig.~\ref{fig1}). This effect being linked to
electron-electron interactions \cite{kravchenko,efros}, its asymmetry
about $\nu=1$ should reflect the electron-hole asymmetry of many-body
interactions in the lowest spin-split Landau level. As the magnetic
field is increased, the experimental magnetocapacitance becomes more
symmetric (see Fig.~\ref{fig4}) and, hence, the electron-hole
symmetry sets in. To this end, it is tempting to give such a
qualitative account of the observed dependence of the spin gap on
perpendicular magnetic field, at least, for the lowest filling
factor. However, there exists a caveat that the exponent for
$\Delta(B_\perp)$ can be expected to decrease with increasing
magnetic field, which is not confirmed by the experimental data.

\section{CONCLUSION}

In summary, we have performed measurements of the chemical potential
jump across the $\nu=1$ and $\nu=3$ many-body enhanced spin gap in
the 2D electron system of GaAs/AlGaAs single heterojunctions in weak
magnetic fields. The increase of the gap with parallel magnetic field
corresponds to a single spin flip for $\nu=1$ and 3 in the range of
magnetic fields studied, down to $B_\perp\approx2$~T. This finding is
in disagreement with the concept of skyrmions and shows that evidence
for skyrmions should be sought in the region of yet lower Zeeman
energies and that results of indirect studies of the spin gap should
be treated with care.

\acknowledgments

We gratefully acknowledge discussions with V.~T. Dolgopolov, V.~I.
Falko, S.~V. Iordanskii, and S.~V. Kravchenko. We would like to thank
P. Pingue, S. Roddaro, and  C. Pascual Garcia for help with the
sample processing. This work was supported by the RFBR, the Programme
``The State Support of Leading Scientific Schools'', and the FIRB
project ``Nanoelectronics'' of the Italian Ministry of Research.


\begin{thebibliography}{apssamp}
\bibitem{ando} T. Ando and Y. Uemura, J.\ Phys.\ Soc.\ Jpn.\ {\bf
37}, 1044 (1974); Yu.~A. Bychkov, S.~V. Iordanskii, and G.~M.
Eliashberg, JETP\ Lett.\ {\bf 33}, 143 (1981); C. Kallin and B.~I.
Halperin, Phys.\ Rev.\ B\ {\bf 30}, 5655 (1984).
\bibitem{sondhi} S.~L. Sondhi, A. Karlhede, S.~A. Kivelson, and
E.~H. Rezayi, Phys.\ Rev.\ B\ {\bf 47}, 16419 (1993).
\bibitem{usher} A. Usher, R.~J. Nicholas, J.~J. Harris, and C.~T.
Foxon, Phys.\ Rev.\ B\ {\bf 41}, 1129 (1990).
\bibitem{schmeller} A. Schmeller, J.~P. Eisenstein, L.~N. Pfeiffer,
and K.~W. West, Phys.\ Rev.\ Lett.\ {\bf 75}, 4290 (1995).
\bibitem{maude} D.~K. Maude, M. Potemski, J.~C. Portal, M. Henini, L.
Eaves, G. Hill, and M.~A. Pate, Phys.\ Rev.\ Lett.\ {\bf 77}, 4604
(1996).
\bibitem{barrett} S.~E. Barrett, G. Dabbagh, L.~N. Pfeiffer, K.~W.
West, and R. Tycko, Phys.\ Rev.\ Lett.\ {\bf 74}, 5112 (1995).
\bibitem{goldberg} E.~H. Aifer, B.~B. Goldberg, and D.~A. Broido,
Phys.\ Rev.\ Lett.\ {\bf 76}, 680 (1996).
\bibitem{qq} I.~V. Kukushkin, K.~von Klitzing, and K. Eberl, Phys.\
Rev.\ B\ {\bf 55}, 10607 (1997).
\bibitem{zhitomir} V. Zhitomirsky, R. Chughtai, R.~J. Nicholas, and
M. Henini, Semicond.\ Sci.\ Technol.\ {\bf 19}, 252 (2004).
\bibitem{terasawa} D. Terasawa, M. Morino, K. Nakada, S. Kozumi, A.
Sawada, Z.~F. Ezawa, N. Kumada, K. Muraki, T. Saku, and Y. Hirayama,
Physica\ E\ {\bf 22}, 52 (2004).
\bibitem{smith} T.~P. Smith, B.~B. Goldberg, P.~J. Stiles, and M.
Heiblum, Phys.\ Rev.\ B\ {\bf 32}, 2696 (1985); T.~P. Smith III,
W.~I. Wang, and P.~J. Stiles, Phys.\ Rev.\ B\ {\bf 34}, 2995 (1986).
\bibitem{aristov} V.~T. Dolgopolov, A.~A. Shashkin, A.~V. Aristov, D.
Schmerek, W. Hansen, J.~P. Kotthaus, and M. Holland, Phys.\ Rev.\
Lett.\ {\bf 79}, 729 (1997).
\bibitem{gapsi} V.~S. Khrapai, A.~A. Shashkin, and V.~T. Dolgopolov,
Phys.\ Rev.\ B\ {\bf 67}, 113305 (2003); Phys.\ Rev.\ Lett.\ {\bf
91}, 126404 (2003).
\bibitem{dobers} M. Dobers, K. v. Klitzing, and G. Weimann, Phys.\
Rev.\ B\ {\bf 38}, 5453 (1988).
\bibitem{zhu} J. Zhu, H.~L. Stormer, L.~N. Pfeiffer, K.~W. Baldwin,
and K.~W. West, Phys.\ Rev.\ Lett.\ {\bf 90}, 056805 (2003); Y.~W.
Tan, J. Zhu, H.~L. Stormer, L.~N. Pfeiffer, K.~W. Baldwin, and K.~W.
West, Phys.\ Rev.\ Lett.\ {\bf 94}, 016405 (2005).
\bibitem{gumbs} A.~P. Smith, A.~H. MacDonald, and G. Gumbs, Phys.\
Rev.\ B\ {\bf 45}, R8829 (1992).
\bibitem{mihalek} I. Mihalek and H.~A. Fertig, Phys.\ Rev.\ B\ {\bf
62}, 13573 (2000).
\bibitem{falko} V.~I. Falko and S.~V. Iordanskii, cond-mat/0003224.
\bibitem{kravchenko} S.~V. Kravchenko, V.~M. Pudalov, and S.~G.
Semenchinsky, Phys.\ Lett.\ A\ {\bf 141}, 71 (1989); S.~V.
Kravchenko, J.~M. Caulfield, J. Singleton, H. Nielsen, and V.~M.
Pudalov, Phys.\ Rev.\ B\ {\bf 47}, 12961 (1993).
\bibitem{efros} A.~L. Efros, F.~G. Pikus, and V.~G. Burnett, Solid\
State\ Commun.\ {\bf 84}, 91 (1992).
\end{thebibliography}
\end{document}